# Optimal Aggregation of Blocks into Subproblems in Linear-Programs with Block-Diagonal-Structure

Deepak Ponvel Chermakani
deepakc@pmail.ntu.edu.sg  deepakc@e.ntu.edu.sg  deepakc@ed-alumni.net  deepakc@hawaii.edu
deepakc@myfastmail.com  deepakc@usa.com

***Abstract: -*** Wall-clock-time is minimized for a solution to a linear-program with block-diagonal-structure, by decomposing the linear-program into as many small-sized subproblems as possible, each block resulting in a separate subproblem, when the number of available parallel-processing-units is at least equal to the number of blocks. This is not necessarily the case when the parallel processing capability is limited, causing multiple subproblems to be serially solved on the same processing-unit. In such a situation, it might be better to aggregate blocks into larger sized subproblems. The optimal aggregation strategy depends on the computing-platform used, and minimizes the average-case running time for the set of subproblems. We show that optimal aggregation is NP-hard when blocks are of unequal size, and that optimal aggregation can be achieved within polynomial-time when blocks are of equal size.

## 1. Introduction

Many practical planning problems can be formulated as Linear-Programs (LPs) with Block-Diagonal-Structure (BDS) with or without common constraints and common variables. BDS in LPs allows decomposition (by either relaxing the common constraints, or fixing the common variables) and hence allows parallelism (i.e. parallel computation) while solving the subproblems. Every decomposition technique is an iterative procedure that runs until a convergence criterion is reached, where each iteration consists of the parallel solution to the set of subproblems, followed by the solution to the master-program. Here, the master problem is formed by the constraints or variables common to all blocks, while the blocks usually form the subproblems. In each iteration, the subproblems retain their original LP structure with altered coefficients, but the master-program might get a new LP structure (depending on the decomposition technique used, by the addition of new constraints or variables) along with altered coefficients. The underlying equivalence of many decomposition techniques for BDS has been proved [1], for example between Dantzig-Wolfe decomposition on an LP, Benders decomposition on the LP's dual, and a cutting-plane approach to solve its Lagrangian dual.

Parallelism is coarse-grained [7] while executing subproblems, meaning that in this context, a subproblem executes continuously from start to finish on the parallel-processing-unit (PPU) to which it is assigned. By PPU, we mean an independent processing unit, which is usually a core on a machine. If one has a large number of PPUs, at least equal to the number of subproblems, then it makes sense to have as many small-sized subproblems as possible, and allow simultaneous execution of each subproblem on a separate PPU, which has been done in [2][3]. But issues [4][5][6][7] have been reported with decomposition, while attempting to solve too many subproblems in parallel, when the number of PPUs is limited (example, on a four-core machine). Some issues are slow convergence, memory-overhead, and communication-overhead.

1.1 Related work
One branch of research [5][6][8][9] into decomposition techniques for LP, focusses on heuristics for row-permutations and column-permutations on an initial sparse constraint matrix, to yield a matrix with specified number of diagonal blocks. The basic approach of these permutation strategies is to represent the non-zero structure of the constraint matrix as a bipartite graph, and reducing the permutation problem to graph-partitioning by vertex separation. These permutation problems are NP-Hard [6] in general, which is why heuristics have been explored for such permutation. However, these permutation-based approaches have not mentioned the optimal number or size of blocks for the given computing-platform.

A second branch of research [10][11][12][13] has been into the theory of the master program, for example generating more effective cutting-planes in Benders decomposition, or improved column-generation for Dantzig-Wolfe decomposition, or better sub-gradient optimization techniques for Lagrangian relaxation. While this research helps in lowering the number of iterations of decomposition techniques, on any computing-platform, it does not address the question of how to reduce the solution-time of the subproblems in every iteration of any decomposition technique.

A third branch of research [14] focuses on scheduling *n* subproblems (of unequal size) parallely on *m* PPUs (*m*<*n*), to minimize overall solution time. This problem is equivalent to the the well-known NP-hard job-shop-scheduling. Due to its hardness, many heuristics have been proposed. While this research reduces solution-time of subproblems in every iteration of any decomposition technique, it will not perform well in a situation where the number of PPUs is limited, and where the solution-time for an LP on the given computing-platform has a high fixed-component that is independent of LP-size.

A fourth branch of research is into aggregating blocks and treating each aggregate as a separate subproblem. This makes sense when two or more subproblems have been scheduled to execute serially on the same PPU. The idea of aggregation has been done earlier [15] as a part of a heuristic (called the "cascading process") while solving LPs with a staircase structure (similar to BDS, except that adjacent blocks share common variables). Aggregating subproblems in a LP with BDS, has been done in real world applications [2][7]. The research in [2][3] concluded that the greater the number of small-sized subproblems in this BDS, the better is the performance, when the number of available PPUs is large. This is however, not the case when the number of PPUs is limited, as agreed by the authors of [5][6][7]. The authors of [5] also gave the idea of performing row and column permutations on a sparse matrix to get a specified number of blocks, but did not specify a strategy to get the optimum number of blocks catered for a given computing-platform.

Aggregation of subproblems has also been recently studied [23][24][25] under the topic of clustering (or grouping) of scenarios of stochastic LPs, such that each cluster (or group) of scenarios is treated as a separate subproblem. These authors used heuristics such as k-means-clustering to create a partition or covering of the set of scenarios, so as hopefully minimize both the running time of the clustered subproblems and the number of iterations of the decomposition technique. It should be noted however that these heuristics are not likely to succeed on all computing platforms. As we will see later, whether or not two or more scenarios (or blocks of the initial LP constraint matrix) should be grouped under the same subproblem, depends on the nature of the computing platform being used.

1.2  Focus of our research

The focus of our research is on reducing the expected solution-time of subproblems, per iteration of any decomposition technique, when the number of PPUs (i.e. cores) is limited to one. This is with the reasonable assumption that the blocks may be divided among available PPUs, so that every PPU is roughly equally loaded. Figures 1, 2 and 3 illustrate how on a three-core machine, two-core machine, and one-core machine respectively, the master-problem and three subproblems (SPs) are scheduled (red means the core is busy, while yellow means the core is idle). We have not found any strategy in literature for LPs with BDS, catered to reduce the solution-time of subproblems on a given computing-platform, giving consideration to to the fixed-component and size-dependent-component in the solution-time of an LP on that computing-platform. To the best of our knowledge, our research is the first to focus on this issue. We will first describe aggregation of blocks into subproblems as an approach to deal with this issue. We will then explore the complexity of aggregation. For those cases of aggregation that are difficult, we will try to prove hardness with well-established complexity classes. For those cases of aggregation that allow an easy solution, we will develop an efficient approach to aggregate blocks into subproblems, so that the sum of the solution times of the subproblems is minimized on the computing-platform, in every iteration of any decomposition technique. We will test our aggregation approach using a well-known LP with BDS.

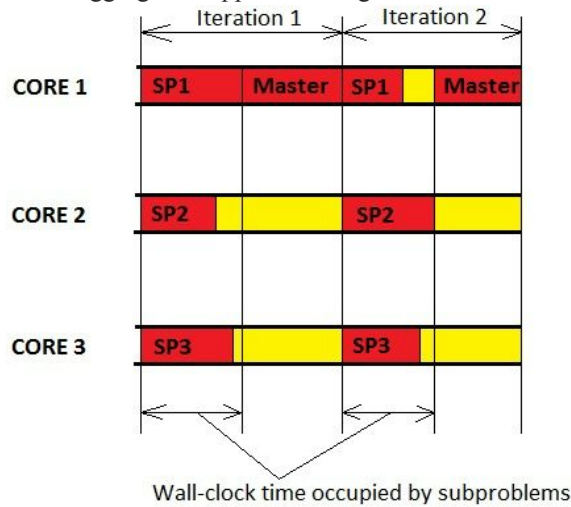

**Figure 1.  Illustration of wall-clock-time spent in solving subproblems on a 3-core machine**

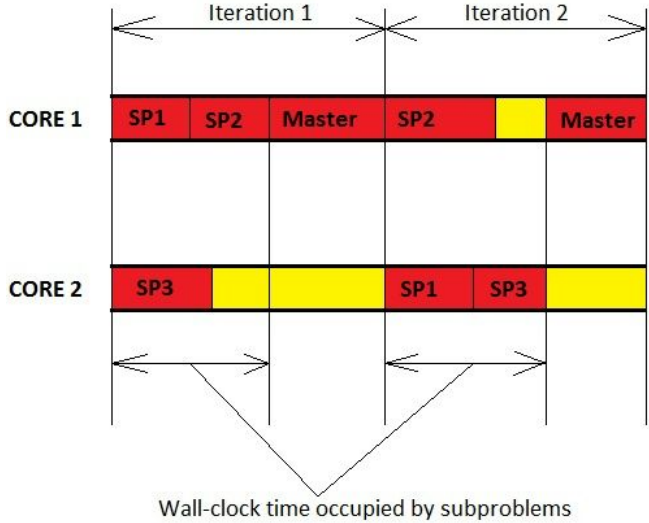

**Figure 2.** Illustration of wall-clock-time spent in solving subproblems on a 2-core machine

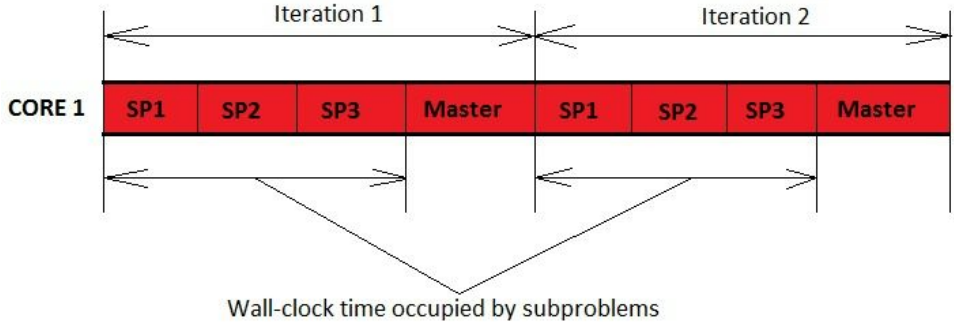

**Figure 3.** Illustration of wall-clock-time spent in solving subproblems on a 1-core machine, which we aim to reduce

## 2. Theory of our approach

In a LP with BDS, aggregation is the grouping of diagonal blocks into one or more groups, so that each group (or aggregate) is treated as a separate subproblem. The purpose of aggregation is to minimize the solution-time of the set of subproblems. In this paper, we aim to minimize solution-time on one PPU, using the assumption that the blocks have been distributed equally among PPUs, giving roughly the same load to each PPU. Figure 4 illustrates aggregation of four blocks into two subproblems, in an LP with BDS with complicating constraints.

We define $P(x)$ to be a polynomial-function specific for a computing-platform, such that for any LP of size $i$, $P(i)$ is the "solution-time" on that computing-platform. If all blocks in the LP with BDS have the same size $s$, $P(si)$ can also be defined as the "solution-time" of a subproblem with $i$ blocks. In this section, by "solution-time", we do not mean exact solution-time (which depends on LP coefficients), but we mean either expected solution-time or worst-case solution-time (whichever is desired to be optimized), over all possible LP coefficients in "desired ranges". If nothing is known about the type of LP that the computing-platform will frequently deal with, the "desired ranges" can simply mean all integers whose magnitudes are within some "large positive integer". This "large positive integer" can be treated as a specified bound or the largest possible integer that can be represented on the computing-platform. $P(x)$ can be treated as a polynomial-function, since the memory of any computing-platform is finite, and it is well-known that a set of $d$ points can be connected by a polynomial of degree $d$.

We illustrate the motivation for aggregating blocks into subproblems with an example. Consider two blocks of sizes $x_1$ and $x_2$, scheduled to run serially on a PPU. Total solution time of subproblems, by not aggregating, is equal to $(P(x_1)+P(x_2))$. Total solution time of subproblems, by aggregating, is equal to $P(x_1 + x_2)$. Hence, whether one should aggregate or not aggregate, depends on whether $P(x_1 + x_2)$ is lesser or greater than $(P(x_1)+P(x_2))$. It is easy to see that $P(x_1 + x_2)$ would be lesser than $(P(x_1)+P(x_2))$ if $P(x)$ has a very large constant term (or fixed component) and if $x_1$ and $x_2$ are both small. In the next section, we give a precise definition to our problem of optimal aggregation of blocks into subproblems.

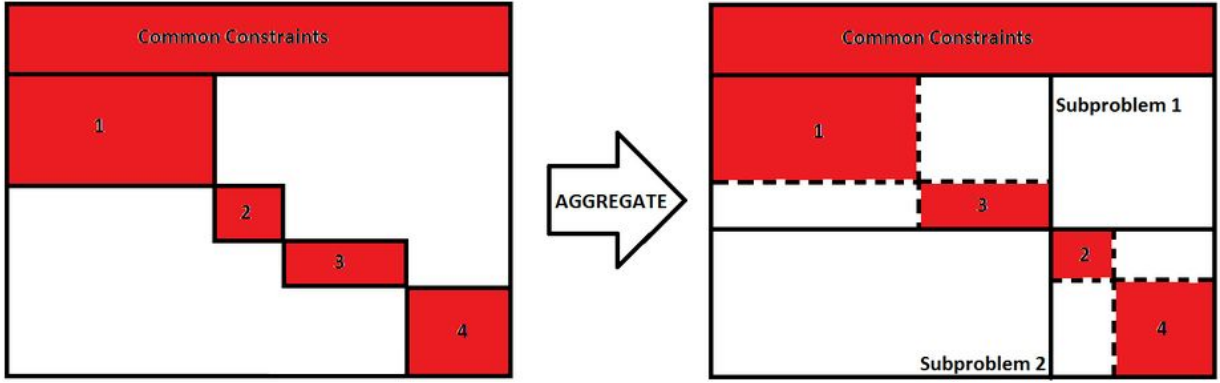

**Figure 4.  Illustration of aggregation**

2.1   Rigorous definition for the problem of optimal aggregation

Given $k$ diagonal blocks of sizes $\{x_1, x_2, ..., x_k\}$ and denote $n = x_1 + x_2 + ... + x_k$. The aim of aggregation is to obtain $k$ subproblems of sizes $\{y_1, y_2, ..., y_k\}$, where every block is assigned to exactly one subproblem, so that the sum of expected solution-times of the subproblems is minimized. Note that some subproblems might be of zero-size. Let $b_{ij}$ be a binary variable that is 1, if and only if, block $i$ is assigned to subproblem $j$. Let $y_j$ be a non-negative real variable depicting the size of subproblem $j$. Let $z_j$ be a binary variable that is 1, if and only if, $y_j$ is positive. We can now define our aggregation problem as the following non-linear-integer-program:

Minimize:
$$\sum_{j=1}^{k} ( P(y_j) - P(0)(1-z_j) )$$

Subject to:
$$\sum_{j=1}^{k} b_{ij} = 1 \text{ , for each integer } i \text{ in } [1,k]$$
$$y_j = \sum_{i=1}^{k} ( x_i b_{ij} ) \text{ , for each integer } j \text{ in } [1,k]$$
$$z_j \leq y_j \leq n z_j \text{ , for each integer } j \text{ in } [1,k]$$
$b_{ij} \in \{0,1\}$ and $z_j \in \{0,1\}$ are binary variables, and $y_j$ is a real variable for all integers $i$ and $j$ in $[1,k]$

In Section 2.2, we show that it is NP-Hard to optimally aggregate blocks of unequal size, into subproblems. In Section 2.3, we show that one can, within polynomial time, optimally aggregate blocks of equal size, into subproblems.

2.2   Optimally aggregating blocks of unequal size into subproblems: An NP-hard problem

**Theorem 1:   Optimally aggregating blocks of unequal size, into subproblems, is strongly NP-Hard**
**Proof:**   We proceed by showing that any instance of a strongly NP-Complete problem can be converted into the question of whether or not the optimal objective value of our aggregation problem can be equal to some value, within polynomial time. The strongly NP-Complete problem we use is 3-PARTITION, one of whose versions is to decide whether or not a given set $X$ of positive integers can be partitioned into three subsets, such that the sum of elements in every subset is the same. Create an instance of our aggregation problem with a set of block sizes that is equal to the set of positive integers $X = \{x_1, x_2, ..., x_k\}$ from any instance of 3-PARTITION. Define $n = \sum_{i=1}^{k} x_i$. Let our growth polynomial be given by $P(x) = a + x^2$, assuming that one can build a computing-platform with such a growth function in $O(Q(log(a)))$ time, where $Q(log(a))$ is a polynomial function of $log(a)$. It is clear that if $a$ is very large, we would aggregate all blocks into a single subproblem of size $n$. It is also clear that if $a$ is 0, we would leave all $k$ blocks without any aggregation. If $a$ is varied between 0 and the very large value, the optimal number of subproblems will vary from $k$ to 1 (need not vary continuously among these integers). Due to the convexity of $P(x)$, for every value of the optimal number of subproblems, the optimal assignment of blocks to subproblems

would occur when all subproblems have the same size, if it is possible. We now consider two cases:

CASE 1 (There exists three partitions of $X$, such the the sum of integers in each partition is equal to $(n/3)$). In this case, we first choose $a$ such that the optimal aggregation strategy is to have 3 subproblems, each of size $(n/3)$. By Theorem 2, this aggregation strategy is optimal when $P'(n/3) = (P(n/3) / (n/3))$, which happens when $a = n^2/9$. The optimal objective value of our aggregation problem in CASE 1 is thus equal to $(3(a + (n/3)^2)) = (2n^2/3)$.

CASE 2 (There do not exist three partitions of $X$, such the the sum of integers in each partition is equal to $(n/3)$). Set $a = n^2/9$. Now, the optimal aggregation strategy can have $h$ subproblems of non-zero size, where $h$ is some integer in $[1,k]$. Denote the sizes of the $h$ subproblems as $\{y_1, y_2, ..., y_h\}$. It is impossible for $y_1 = y_2 = ... = y_h = (n/3)$, due to the condition imposed by CASE 2. Also note that the least value of $(P(x)/x)$ occurs when $x = (n/3)$. Therefore, $((\sum_{i=1}^{h} P(y_i))/n) = (\sum_{i=1}^{h} P(y_i) / \sum_{i=1}^{h} y_i) = (\sum_{i=1}^{h} (y_i P(y_i)/y_i) / \sum_{i=1}^{h} y_i)$ is the weighted average of $h$ values, each of which is greater than or equal to $(P(n/3)/(n/3))$, and one of which is strictly greater than $(P(n/3)/(n/3))$. This is due to the convexity of $P(x)$. The optimal objective value of our aggregation problem in CASE 2 is thus equal to $((2n^2/3) + \Delta)$, where $\Delta > 0$.

From CASE 1 and CASE 2, by setting $a = (n^2/9)$, the answer to the 3-PARTITION problem is YES, if and only if, the optimal objective to the aggregation problem is equal to $(2n^2/3)$. We have reduced 3-PARTITION into the decision version of our aggregation problem, within polynomial time, showing its strong NP-Hardness. **Hence Proved**

### 2.3 Optimally aggregating blocks of equal size, into subproblems, within polynomial time

Denote the size of each block as $s$, the sum of sizes of blocks per PPU as $n$, and hence the total number of blocks per PPU as $(n/s)$. We first give Theorem 2 for obtaining the optimal subproblem size under the assumption of continuity.

**Theorem 2:** The optimal size (denoted as $x_{opt}$) for a subproblem, assuming $x_{opt}$ and $(n/x_{opt})$ are both integers, is the value of $x$ for which the least-sloped tangent to $P(x)$ meets the origin.

**Proof:** One possible candidate for the optimal size, will involve setting all subproblems to have the same size $x_{opt}$, since if we are able to find the optimal size for one subproblem, then all subproblems can be set to that same size, so that overall solution time is minimized for all subproblems. The sum of sizes of all subproblems is $n$. Assuming $x$ to be an integer, denote the size of each subproblem as $x$. So the solution time for each subproblem is $P(x)$, and the number of subproblems is $(n/x)$, where assume $(n/x)$ to also be an integer. Denoting $T$ to be the total solution time for all subproblems on an IPU, we can say that $T = (n/x) P(x) = n(P(x)/x)$. Since $n$ is fixed, the global minimum of $T$ has the least value of $(P(x)/x)$. The set of stationary points of $T$ are obtained by setting $dT/dx = 0$ giving $P'(x) = (P(x)/x)$, where $P'(x)$ represents $d(P(x))/dx$. This means $x_{opt}$ is the value of $x$, at which the least-sloped tangent to $P(x)$ meets the origin. **Hence Proved**

If $x_{opt}$ is an integer multiple of $s$, and if $n$ is an integer multiple of $x_{opt}$, we are done by setting the sizes of all subproblems to $x_{opt}$. However, if $x_{opt}$ is not an integer multiple of $s$, or, if $n$ is not an integer multiple of $x_{opt}$, we may follow a Dynamic Programming (DP) approach that gives an optimal aggregation strategy in $O((n/s)^2)$ time. We give Theorem 3, which converts our problem of optimal aggregation of blocks of equal size, into a DP problem, and gives the DP approach.

**Theorem 3:** Let $P(x)$ be an arbitrary function with domain $[1, n]$. Let all $(n/s)$ blocks be of equal size $s$. The optimal solution to our aggregation problem can be obtained in $O((n/s)^2)$ time using a Dynamic Programming approach.

**Proof:** Denoting $x_i$ as the number of subproblems to be formed of size $si$ (i.e. with $i$ blocks), our aggregation problem is equivalent to the following Integer Linear Program (ILP):

Minimize:
$$\sum_{i=1}^{(n/s)} x_i P(is)$$
Subject to:
$$\sum_{i=1}^{(n/s)} (ix_i) = (n/s)$$
$x_i$ is an integer variable in $[0, (n/s)]$, for all integers $i$ in $[1, (n/s)]$

The above ILP formulation is the well-known Unbounded-Knapsack problem with equality-constraint, for which the following pseudo-polynomial DP algorithm is also well-known [22], by converting into a shortest path problem:

1) Define $(n/s)$ states $\{f_0, f_1, f_2, ... f_i ..., f_{(n/s)}\}$, where $f_i$ denotes the optimal solution time of $i$ blocks, each of size $s$
2) For all $i > j$, $c_{ij} = P((i-j)s)$
3) $f_0 = 0$

4) $f_i = \min_{j \in \{i-1, i-2, \ldots, 2, 1, 0\}} (c_{ij} + f_j)$, for all integers $i$ in $[1, (n/s)]$
5) Initialize $x_i = 0$ for all integers $i$ in $[1, (n/s)]$
6) Retrace the path backwards starting from $f_{(n/s)}$, where from each state $i$, we choose to go to state $t$, where $t = \text{argmin}_{j \in \{i-1, i-2, \ldots, 2, 1, 0\}} (c_{ij} + f_j)$, at the same time updating $x_{i-t} = x_{i-t} + 1$

The complexity is clearly $O(1 + 2 + 3 + \ldots + ((n/s)-1) + (n/s)) = O((n/s)^2)$. **Hence Proved**

When $P(x)$ is known to be convex or concave, we can use certain properties to obtain algorithms that are less complex than $O((n/s)^2)$. Theorems 4 and 5 describe these properties for convex and concave functions respectively.

**Theorem 4:** Let $P(x)$ be a convex function in the domain $[sr_1, sr_2]$, where $r_1$ and $r_2$ are positive integers. Let all $(n/s)$ blocks be of equal size $s$. If a feasible solution exists to the aggregation problem, there exists an optimal solution that consists of not more than two unique subproblem sizes in $[sr_1, sr_2]$, and these two sizes differ by not more than $s$.
**Proof:** Assume there exists an optimal solution having more than two unique subproblem sizes. Iteratively subtract $s$ from the size of the largest subproblem, and add $s$ to the size of the smallest subproblem, causing overall solution time to either reduce or remain the same (due to convexity property). In this case, since our solution was assumed to be optimal, the overall solution time remains the same. We eventually get no more than two unique subproblem sizes, which differ by not more than $s$. **Hence Proved**

**Theorem 5:** Let $P(x)$ be a concave function in the domain $[sr_1, sr_2]$, where $r_1$ and $r_2$ are positive integers. Let all $(n/s)$ blocks be of equal size $s$. If a feasible solution exists to the aggregation problem, there exists an optimal solution that consists of not more than three unique subproblem sizes, belonging to the set $\{sr_1, sr, sr_2\}$, where $r_1 < r < r_2$, and $r$ is some integer.
**Proof:** Assume that an optimal solution has four or more unique subproblem sizes in $[sr_1, sr_2]$. Iteratively consider any two of these subproblems, whose sizes are neither equal to $sr_1$ nor equal to $sr_2$, and subtract $s$ from the size of the smaller subproblem and add $s$ to the size of the larger subproblem, causing overall solution time to either reduce or remain the same (due to concavity property), until one of their sizes is equal to either $sr_1$ or $sr_2$. In this case, since our solution was assumed to be optimal, the overall solution time remains the same. We eventually get not more than three unique subproblem sizes, belonging to the set $\{sr_1, sr, sr_2\}$, where $r_1 < r < r_2$, and $r$ is some integer. **Hence Proved**

We now use the properties described in Theorem 4 and Theorem 5, to develop approaches for convex functions (see Theorem 6) and for concave functions (see Theorem 7) that are less complex than $O((n/s)^2)$.

**Theorem 6:** Let $P(x)$ be a convex function with domain $[1, n]$. Let all $(n/s)$ blocks be of equal size $s$. Let $rs$ be the smallest integer multiple of $s$ greater than $x_{opt}$. The optimal solution to the aggregation problem, can be obtained in:
1) $O(n/s)$ time for all $(n/s) \leq (r^2 - r - 1)$
2) $O(1)$ time for all $(n/s) > (r^2 - r - 1)$

**Proof:** We consider three cases:
CASE 1 ($x_{opt} \leq 1$): This is a trivial case in which it is optimal is to have no-aggregation (i.e. each block becomes a separate subproblem). Solution time is $O(1)$.
CASE 2 ($x_{opt} \geq n$): This is a trivial case in which it is optimal to have full-aggregation (i.e. all blocks are combined into a single subproblem). Solution time is $O(1)$.
CASE 3 ($1 < x_{opt} < n$): Since $P(x)$ is convex, Theorem 4 applies. In the optimal solution, denote the two unique sizes of the smaller and larger subproblems as $(ks)$ and $((k+1)s)$ respectively, where $k$ is an integer in $[1, ((n/s)-1)]$. Hence, for each of the $((n/s)-1)$ values of $k$, we can obtain a separate ILP:

Minimize:
$$x_1 P(ks) + x_2 P((k+1)s)$$
Subject to:
$$x_1(k)s + x_2(k+1)s = n$$
Each of $\{x_1, x_2\}$ is a non-negative integer variable $\leq (n/s)$

For each value of $k$, the above ILP can be solved by an algorithm [16][17] within constant time, since the number of variables and constraints is constant. We choose the solution with minimum objective value from amongst the $((n/s)-1)$ ILPs. Overall complexity is thus $O(n/s)$.

Due to convexity of $P(x)$, the two integer multiples of $s$, with the least values of $(P(x)/x)$, are among the set $\{(r-2)s, (r-1)s, rs, (r+1)s\}$. Applying the formula [21] of the Frobenius number of two coins with denominations $rs$ and $(r+1)s$, the equation:

$x_1 s(r)+x_2 s(r+1)=n$, or equivalently: $x_1(r)+x_2(r+1)=(n/s)$, has a solution for all values of $(n/s) > (r(r+1)-(r+1)-r) = (r^2 - r - 1)$. The value of $(r^2 - r - 1)$ attains a minimum at $r=0.5$, meaning that $(r^2 - r - 1)$ increases for all $r>1$. This also means that for all integers $k$ in $[1, r]$, the general equation $(x_1(k)+x_2(k+1)=(n/s))$ has a solution if $(n/s) > (r^2 - r - 1)$. Using Theorem 4 and for minimizing the weighted average of $(P(x)/x)$, it suffices to consider only three ILPs, the first ILP being with $k=(r-2)$, the second ILP being with $k=(r-1)$, and the third ILP being with $k=r$. Hence, overall complexity is O(1) for all $(n/s) > (r^2 - r - 1)$. **Hence Proved**

**Theorem 7:** Let $P(x)$ be a concave function with domain $[1, n]$. Let all $(n/s)$ blocks be of equal size $s$. The optimal solution to the aggregation problem, can be obtained in $O((n/s))$ time

**Proof:** Since $P(x)$ is concave, Theorem 5 applies. In the optimal solution, denote the unique size of the subproblem that is between the smallest and largest integer multiples of $s$, as $(ks)$, where $k$ is some integer in $[2, ((n/s)-1)]$. Hence, for each of the $((n/s)-2)$ values of $k$, we can obtain a separate ILP:

Minimize:
$$x_{SMALLEST} P(k_{SMALLEST}\, s) + x P(ks) + x_{LARGEST} P(k_{LARGEST}\, s)$$
Subject to:
$$x_{SMALLEST}(k_{SMALLEST}\, s) + x(ks) + x_{LARGEST}(k_{LARGEST}\, s) = n$$
$$k_{SMALLEST} = 1$$
$$k_{LARGEST} = (n/s)$$
Each of $\{x_{SMALLEST}, x, x_{LARGEST}\}$ is a non-negative integer variable $\leq (n/s)$

For each value of $k$, the above ILP can be solved by an algorithm [16][17] within constant time, since the number of variables and constraints is constant. We choose the solution with minimum objective value from amongst the $((n/s)-2)$ ILPs. Overall complexity is thus $O(n/s)$. **Hence Proved**

We now give Theorem 8 that describes the optimal aggregation approach if $P(x)$ is linear.

**Theorem 8:** Let $P(x)$ be a linear function with domain $[1, n]$. Let all $(n/s)$ blocks be of equal size $s$. The optimal solution to the aggregation problem, can be obtained in $O(1)$ time

**Proof:** Let $P(x) = a+bx$. The solution-time of all subproblems $= ka+bn$, where $k$ is the number of subproblems formed, and hence $k$ is an integer in $[a, na]$. If $a \geq 0$, to minimize the value of $ka+bn$, we minimize the number of subproblems formed (i.e. aggregate all blocks into a single subproblem). If $a < 0$, to minimize the value of $ka+bn$, we maximize the number of subproblems formed (i.e. treat each block as a separate subproblem). The optimal aggregation strategy is thus devised in $O(1)$ time. **Hence Proved**

2.4  The overall approach

Our aggregation strategy for $(n/s)$ blocks each of size $s$, is as follows:
1) Obtain $P(si)$ for all integers $i$ in $[1,(n/s)]$. This is a one-time task for the given computing-platform.
2) If $P(si)$ is linear, use the $O(1)$ approach of Theorem 8 and exit.
3) If $P(si)$ is either strictly-convex, use the $O(n/s)$ approach of Theorem 6, and exit.
4) If $P(si)$ is either strictly-concave, use the $O(n/s)$ approach of Theorem 7, and exit.
5) Since $P(x)$ is non-convex, use the $O((n/s)^2)$ DP approach of Theorem 3.
6) Exit.

## 3. Conclusion

We proposed aggregation of blocks into subproblems, as a strategy to minimize average-case solution-time of an LP with BDS, on a computing-platform with a limited number of PPUs (cores in this case, since we consider a single machine). Assuming that the blocks can be roughly equally divided among the cores of the machine, the focus of our work was optimal aggregation on a single PPU. This makes wall-clock-time equal to overall-cpu-time for solving the subproblems, assuming no interference from foreign processes.

We showed that though optimal aggregation is NP-hard in general, blocks of equal size can be optimally aggregated in polynomial-time. We defined a polynomial-growth-function $P(x)$ that is specific for the computing platform. $P(x)$ is the average-case solution-time (over all coefficients within desired ranges) for an LP of size $x$ on that computing-platform. With an LP having blocks of equal size $s$, $P(si)$ is the average-case solution-time of $i$ blocks each of size $s$, on the given computing-platform. When all ($n/s$) blocks in the LP with BDS, are of the same size $s$, the DP algorithm gives the optimal aggregation strategy in $O((n/s)^2)$ time. When $P(x)$ is either convex or concave, the optimal aggregation strategy can be obtained in $O(n/s)$ time. When $P(x)$ is linear, optimal aggregation can be achieved in $O(1)$ time.

We hope our results will be useful to researchers with limited budgets, to applications in resource constrained computing environments, where it becomes necessary to use hardware machines with limited amounts of cache and limited number of PPUs, and also to HPC systems solving large LPs with a high ratio of subproblems to cores.

**About the author**
I, Deepak Ponvel Chermakani, wrote this paper out of my own interest and initiative, during my spare time. I am currently a graduate student at the University of Hawaii at Manoa USA (www.hawaii.edu). In Sep 2010, I completed a fulltime one year Master Degree in Operations Research with Computational Optimization, from University of Edinburgh UK (www.ed.ac.uk). In Jul 2003, I completed a fulltime four year Bachelor Degree in Electrical and Electronic Engineering, from Nanyang Technological University Singapore (www.ntu.edu.sg). In Jul 1999, I completed fulltime high schooling from National Public School in Bangalore in India.